\documentclass{article}
\usepackage[utf8]{inputenc}
\usepackage{color}
\usepackage{authblk}
\usepackage{geometry}                
\usepackage{graphicx}
\usepackage{amssymb}
\usepackage{epstopdf}
\usepackage{soul}
\usepackage{amsmath,amsbsy}
\usepackage[shortlabels]{enumitem}
\newcommand{\MM}[1]{\textcolor{black}{{#1}}}
\newcommand{\NM}[1]{\textcolor{black}{{#1}}}

\newcommand{\RR}[1]{\textcolor{black}{{#1}}}
\newcommand {\bs} {\mbox{\boldmath $s$}}

\newcommand {\bn} {\mbox{\boldmath $n$}}
\newcommand*\Bell{\ensuremath{\boldsymbol\ell}}
\newcommand*\Btheta{\ensuremath{\boldsymbol\theta}}
\newcommand*\Bepsilon{\ensuremath{\boldsymbol\epsilon}}
\newcommand*\Bphi{\ensuremath{\boldsymbol\phi}}
\newcommand*\BPhi{\ensuremath{\boldsymbol\Phi}}
\newcommand*\Bsigma{\ensuremath{\boldsymbol\sigma}}
\newcommand {\cF} {\mathcal{F}}
\newcommand {\cS} {\mathcal{S}}
\newcommand {\cR} {\mathcal{R}}

\begin{document}

\title{\textbf{Optimal Work Extraction and the\\ Minimum Description Length Principle }}
\author[1]{\MM{L\'eo Touzo}}
\author[2]{Matteo Marsili}
\author[3]{Neri Merhav}
\author[2]{\'Edgar Rold\'an}
\affil[1]{{\small Department of Physics, \'Ecole Normale Sup\'erieure, 24 rue Lhomond, 75005 Paris, France}}
\affil[2]{{\small The Abdus Salam International Centre for Theoretical Physics, Strada Costiera 11, 34151 Trieste, Italy}}
\affil[3]{{\small The Andrew and Erna Viterbi Faculty of Electrical Engineering, Technion - Israel Institute of Technology
Technion City, Haifa 3200003, Israel}}
\date{}

\maketitle

\begin{abstract}
We discuss work extraction from  classical information engines (e.g., Szil\'ard) with $N$-particles, $q$ partitions, and initial arbitrary non-equilibrium states. 
In particular, we focus on their {\em optimal} behaviour, which includes the measurement of a set of quantities $\BPhi$
with a feedback protocol that extracts the maximal average amount of work. 
We show that the optimal non-equilibrium state to which the engine should be driven before the measurement is given by the normalised maximum-likelihood  probability distribution of a statistical model that admits $\BPhi$ as sufficient statistics. Furthermore, 
we show that the minimax universal code redundancy $\mathcal{R}^{\star}$ associated to this model, provides an upper bound to the 
work that the demon can extract on average from the cycle, in units of $k_{\rm B}T$.
We also find that, in the limit of $N$ large, the maximum average extracted work 
cannot exceed $H[\BPhi]/2$, i.e. one half times the Shannon entropy of the 
measurement.
 Our results   establish a  connection between optimal work extraction in stochastic thermodynamics and optimal  universal data compression, providing design principles for optimal information engines. In particular, they suggest that: (i) optimal coding is  thermodynamically efficient, and (ii) it is essential to drive the system into a critical state in order to
achieve optimal performance.
\end{abstract}

\rightline{In memory of Jorma J. Rissanen,}
\rightline{the father of the minimum description length principle,}
\rightline{and a great inspiration to many of us, for years to come.}

\section{Introduction and statement of the main results}

Bits are emerging as a universal currency in a {wide
 variety} of sciences, {ranging} from
information theory and coding theory, to statistical mechanics and thermodynamics~\cite{landauer1961irreversibility,bennett1982thermodynamics,leff2002maxwell,parrondo2015thermodynamics,wolpertbook2019}. In the spirit of Maxwell's demon, Szil\'ard's information engine~\cite{Szilard-german}  is considered by many as the founding stone of information thermodynamics. Szil\'ard's information engine executes a cyclic protocol by using one bit of information of a physical system, yielding to the extraction of $k_{\rm B}T\ln 2$   of energy from a thermal bath at temperature $T$, in apparent contradiction with the second law of thermodynamics~\cite{landauer1961irreversibility, bennett1982thermodynamics,leff2002maxwell}.  Recent progress within the emerging field of stochastic thermodynamics has led to a deeper understanding of the relationship between information and entropy at mesoscopic scales, leading to e.g., the refinement of the second law of thermodynamics for  systems  with feedback control~\cite{sagawa2012thermodynamics}.  These works study fundamental
thermodynamic bounds governing processes that convert information into work and/or heat, and
vice versa, both in theory~\cite{sagawa2009minimal,SagawaUedaFeedback12,mandal2012work,Horowitz2011,kim2011quantum,kim2011information, mandal2014, barato2014unifying, lathouwers2017memory,bengtsson2018quantum,chetrite2019information,wolpertbook2019,still2020thermodynamic}, and experiment~\cite{toyabe2010experimental,berut2012experimental,roldan2014universal,koski2014experimental, exp-landauer2014,gavrilov2016erasure,gavrilov2017direct,paneru2018lossless,ribezzi2019large}.

Since Maxwell's and Carnot's~\cite{carnot1978reflexions} pioneering work, optimizing the performance  of heat and information engines has remained one of the  key goals of thermodynamics. Substantial  progress  towards  efficient rectification of  fluctuations by mesoscopic machines has been achieved during the last decades within framework of  stochastic thermodynamics~\cite{sekimoto2010stochastic,seifert2012stochastic}. Important examples include    finite-time protocols that enable optimization of different thermodynamic quantities, such as: (i)  minimizing the average heat dissipation and work  needed to drive a system out of equilibrium~\cite{sekimoto1997complementarity,schmiedl2007optimal,then2008computing,sivak2012thermodynamic,bonancca2014optimal}; (ii)  minimizing the uncertainty (e.g., Fano factor) of the work done on a generic non-equilibrium process~\cite{solon2018phase}; (iii) "one shot" optimization leading to work extraction in every realization above a prescribed threshold value~\cite{aaberg2013truly,halpern2015introducing}; (iv)   enforcing a desired value for the probability to extract work by an arbitrarily large value above  the free energy change~\cite{cavina2016optimal,maillet2019optimal}; and (v) reversible feedback-control protocols which ensure an average work dissipation equal to the information gain  during a non-equilibrium process ~\cite{horowitz2011thermodynamic,horowitz2011designing,granger2016reversible}. For general non-equilibrium processes involving  feedback control (e.g. information engines), it remains yet unclear whether there exist fundamental principles linking optimal work extraction and efficient data compression via e.g., optimal coding of the information used in feedback control. Until now, few insights have been developed yet at the interface between stochastic thermodynamics and coding, such as the relation between work and entropy production with sequence complexity~\cite{roldan2010estimating,roldan2012entropy,merhav2015sequence} and channel capacity~\cite{frishman2020learning}.

\begin{figure}[hb]
\centering
\includegraphics[width=\linewidth]{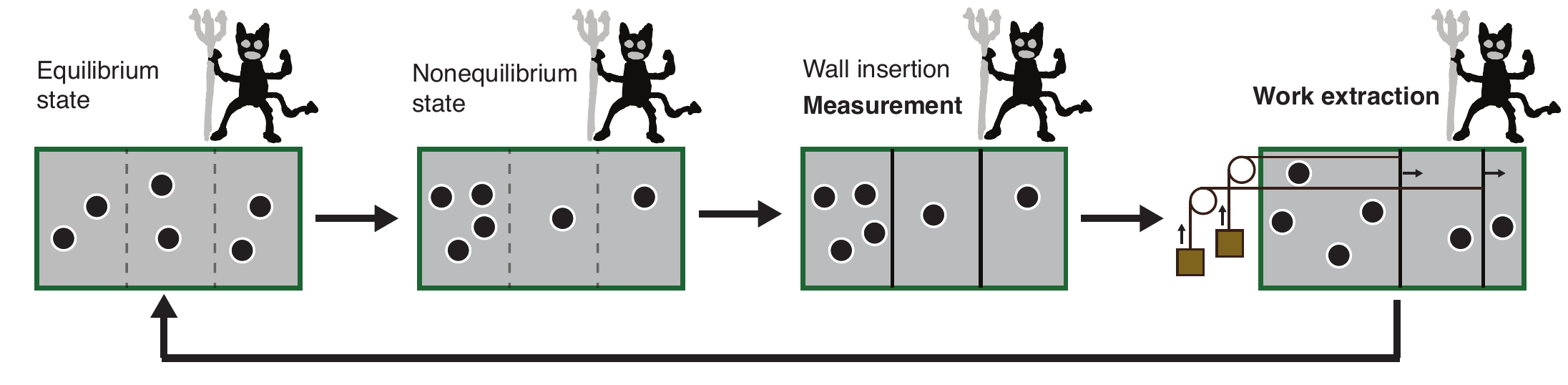}
\caption{\small Illustration of a Szil\'ard engine with $N=6$ particles and $q=3$ partitions. The engine is composed by $N$ non-interacting, distinguishable, classical point particles confined in a box that is in contact with a bath at temperature $T$. 
Initially, the system is in an equilibrium state  with the $N$ particles distributed homogeneously in the box.  Next, 
the system  is driven   to a non-equilibrium state with inhomogeneous spatial distribution. An intelligent ``demon" introduces walls in the box at prescribed locations, and measures the number of particles located in each of the resulting partitions. Using this measurement, the demon is able to lift weights (i.e. to extract work) from the partitions that expand reversibly to a new equilibrium state with equal pressure in each of the partitions. The cycle is then completed by removing the walls and letting the particles relax to equilibrium. Here, we investigate  what is the maximum work that can be extracted on average over many repetitions of the cycle, and what is the non-equilibrium distribution at which the maximum average work is attained. For the Szil\'ard engine in the Figure, the maximum work extracted is given by  $k_{\rm B}T \mathcal{R}^{\star}$, with $\mathcal{R}^*=\ln(3167/324)\simeq 2.28$ the \NM{minimax redundancy}, following Eq.~\eqref{StocCompl0}.}
\label{fig:1}
\end{figure}

In this  paper, we consider  classical many-particle  information engines  that extract work 
from a cycle in which a {certain} microscopic quantity is measured. In particular, we first consider a variant of the $N$-particle Szil\'ard engine with $q$ partitions in which the system is  prepared in a non-equilibrium state before the measurement, see Fig.~\ref{fig:1} for an illustration. An important question is the following: 
what is the maximal {amount of } work that the engine can extract on average from a measurement, and for which initial non-equilibrium distribution?
We limit our discussion {to} work extraction protocols that, as in the original Szil\'ard engine, 
rely on reversible expansion of the partitions after the measurement. 
Our main result is that, for this class of systems, the average work that can be extracted cannot exceed the \NM{minimax universal code redundancy} of the statistical model that admits the measured quantities as sufficient statistics, according to Minimum Description Length
{(MDL) principle} \cite{grunwald2007minimum}, i.e.{,}
\begin{equation}
\label{main_result}
\langle W_{\rm ext}\rangle \le k_{\rm B} T\, \mathcal{R}^{\star},
\end{equation}
with equality holding under optimal protocols. 
Here{,} $W_{\rm ext}$ {denotes} the work extracted in a cycle of the engine, which includes  the
work needed to prepare the non-equilibrium state and the one extracted from the 
feedback protocol after the measurement. The \NM{minimax universal code redundancy{,} $\mathcal{R}^{\star}${,} is a measure of model complexity. Loosely speaking, it}
quantifies the 
amount of information (measured in nats) that the measurement provides about the
parameters of the model. 

Equation (\ref{main_result}) relates a  thermodynamic quantity, on the left hand side,
to a central concept in {source} coding theory, on the
right hand side.
We explore this connection  via a formal equivalence between a 
system of classical non-interacting particles and optimal {universal} coding of a sample of 
independent observations. In this equivalence, the maximization of the extracted work corresponds 
to the maximization of the likelihood.  Indeed, we show that the non-equilibrium state to which the engine should be
driven before the measurement corresponds to the {\em
normalised maximum likelihood} {(NML) probability distribution}, which is
the coding {distribution} that provides optimal{, universal data}
compression  according to the  MDL {principle} \cite{grunwald2007minimum}. 
{This code has} been shown to have critical
statistical{--}mechanical properties in a precise sense~\cite{CuberoMDL}, showing once again the relevance of criticality in information processing systems~\cite{cubero2019statistical}.  For the $N$-particle Szil\'ard engine with $q$ partitions we provide explicit expressions for $\mathcal{R}^*$ $-$Eq.~(\ref{StocCompl0})$-$ and for the optimal non-equilibrium distribution $-$Eq.~(\ref{nmlmn})$-$ as well as numerical values in specific cases (Fig. \ref{fig:szexamples}).

Notably, we show that the connection between optimal work and optimal coding generalizes beyond the Szil\'ard engine to a broad class of models, provided that 
the equilibrium state of the engine is uniquely identified by the measurement. 
Thus, the measured quantities  should be sufficient statistics of the model describing the equilibrium state. \RR{This condition, and the Pitman-Koopman-Darmois theorem \cite{koopman1936distributions}, requires the equilibrium state to have the form of an exponential family.}

Our results suggest design principles of optimal information
engines, such as: (i) how the equilibrium state should be defined, 
depending on the quantity that is measured, (ii) what is the non-equilibrium state 
to which the system should be driven; and (iii) the fundamental limits on the  work that can be extracted. 
\MM{With respect to the latter, an interesting issue is how close our bound gets to the fundamental limit~\cite{sagawa2012thermodynamics} $\langle W_{\rm ext}\rangle \le k_{\rm B} T\,H[\BPhi]$, where $H[\BPhi]$ is the entropy of the measurement $\BPhi$. We find that, in the limit $N\to\infty$ the protocol studied here comes half-way to the upper bound, in the sense that $\langle W_{\rm ext}\rangle / (k_{\rm B} T\,H[\BPhi])$ converges to $1/2$ when $N\to\infty$. This is reminiscent of recent bounds derived on the sensory capacity~\cite{hartich2016sensory}, and on the relevance of sufficient statistics to achieve these bounds~\cite{matsumoto2018role}.}

The rest of the paper is organised as follows: In Section~\ref{sec:ii}, we revisit the
classical Szil\'ard engine with $N$ particles and $q$ partitions~\cite{song2019optimal}, pose our main question, and draw a connection between optimal average work extraction and {the} MDL
{principle}~\cite{grunwald2007minimum}. In Section~\ref{sec:iii}, we map the
Szil\'ard engine to a system  
of non-interacting particles in $q$ energy levels and re-derive this 
connection in a broader framework.
Finally, in Section~\ref{sec:iv}, we discuss the implications and
possible {future} extensions of our work.

\section{Minimum description length and work extraction in the $N$-particle Szil\'ard's engine}
\label{sec:ii}

In this section, we revisit  a variant of the classical Szil\'ard engine modelled by a classical ideal gas  of $N$ particles confined in the $x${-}dimension of a one-dimensional box  of length one and unit area in the perpendicular directions. 
This system has been extensively studied in \cite{song2019optimal}, 
to which we refer for a detailed discussion. Here, we first  
review its main properties and introduce the notation used throughout the paper. Next, we discuss optimal work extraction and relate it to optimal coding of the  information measured by the demon.

\subsection{Definition of the model and notation}

We consider an information engine operating cyclically and consisting of $N\geq 1$ point, non-interacting, particles confined in a box. The system is assumed to be initially in equilibrium with a heat bath at temperature $T$. 
The temperature of the {heat} bath 
is kept constant all along the cycle, hence we take $k_{\rm B} T =1$ henceforth.  
In the initial state, the positions $x_i$ ($i = 1,\dots,N$) of the particles are
dynamical variables spanning the whole box, with a uniform 
distribution. We will denote by $\mathbf{x}=(x_1,\ldots, x_N)$  the vector of the positions of all particles.

\begin{figure}[h!]
\centering
\includegraphics[width=0.8\linewidth]{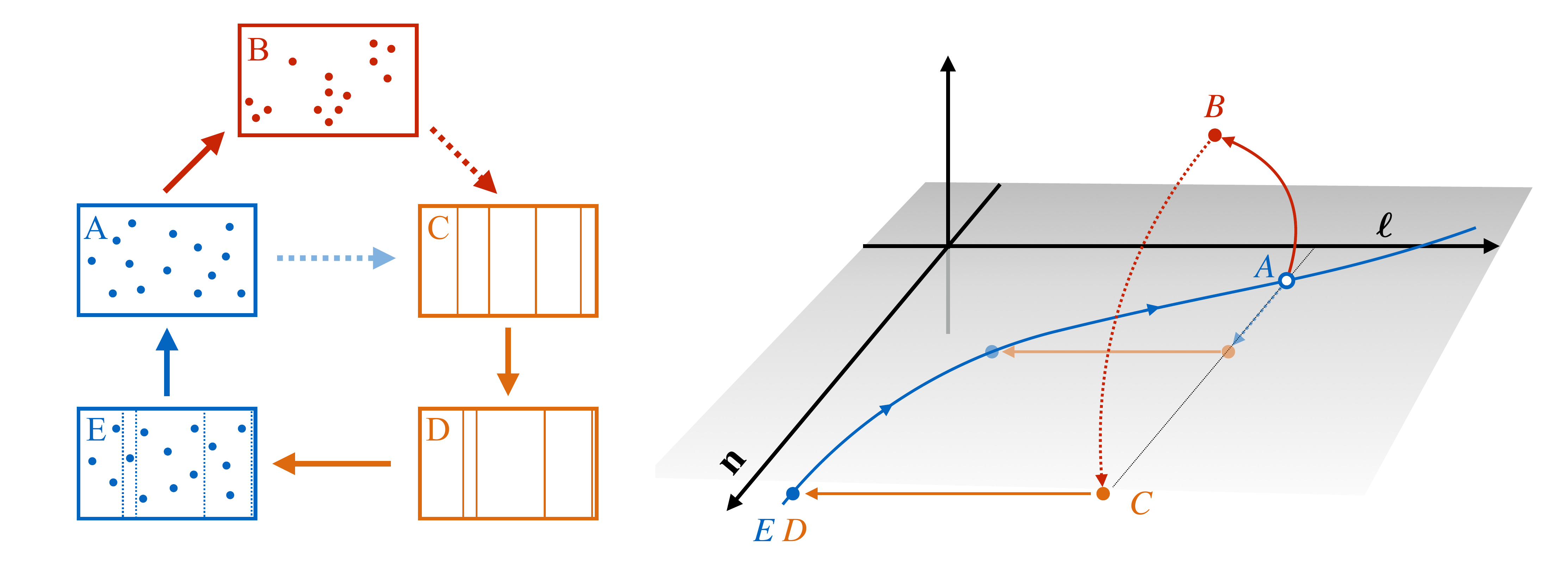}
\caption{\small Schematic representation of the cycle performed by the information engine (left) and pictorial sketch of 
the corresponding transformations in parameter space (right). 
The initial equilibrium state (A) lies on the blue curve that designates the equation of state 
in the (hyper)plane of the parameters $\Bell$ (the partitions' lengths) and the (average of the) sufficient statistics $\mathbf{n}$ (the number of particle in each partition). 
In the preparation phase, the system is driven out of equilibrium 
(red arrow) to a state B (red dot/box). When partitions are
introduced{,} and $\mathbf{n}$ is measured, the system is projected back on the plane to a different point C (dashed red arrow). 
The sufficient statistics are measured and the work extraction protocol is operated (orange arrow) driving the system to state D.
Then the partitions are removed bringing the system back on the equilibrium curve (E). The cycle is completed by
driving back the system to the equilibrium state A. 
In the setting of Ref.~\cite{song2019optimal}, the measurement is taken on the equilibrium state A (light blue arrow) and work is extracted (light orange arrow) from the resulting state.}
\label{FigCycle}
\end{figure}

During the operation of the engine, the box is divided into $q$ partitions. 
 The partitions divide the unit interval into $q$ sub-intervals, $[y_{s-1},y_{s-1}+\ell_s)$,  where $y_0=0$,
$y_{s}=y_{s-1}+\ell_s$ (for all $s=1,\ldots,q$) and $y_{q}=1$. The lengths of the partitions can take arbitrary positive values, consistent with the constraint $\sum_{s=1}^q \ell_s =1$.  We denote by $\Bell=(\ell_1,\dots,\ell_{q})$ the set of lengths of each partition. 
 We now introduce two key quantities that we will use in the following to describe the dynamics of the engine: 
\begin{itemize}
\item The state of the system $\mathbf{s}=(s_1,\dots,s_N)$  is a vector composed of a collection of $N$ coarse-grained integer variables, with  $s_i\in \{1,\dots ,q\}$ denoting the partition in which the $i$-th particle is located. Each particle, $i=1,\dots, N$, is assigned to a partition of the $x$ axis, $s_i=s(x_i)$, where $s(x_i)$  is the smallest 
$s$ for which $x_i>y_s$.
\item The measured quantity $\mathbf{n}=(n_1,\dots,n_q)$ is a vector composed of $q$ integer variables $n_s\in \{0,\dots ,N\}$ denoting the number of particles located in the $s$-th partition. Each partition, $s=1,\dots, q$ contains   $n_s= \sum_{i=1}^{N} \delta_{s_i,s}$ particles, with the total number of particles  $\sum_{s=1}^q n_s=N$ conserved during the cycle. We henceforth refer to $n_s$ ($s=1,\dots,q$) as occupation numbers. 
 \end{itemize} 
We  assume that $\Bell$ and $N$ are known, while $\mathbf{n}$ is obtained
through a measurement, as explained below\footnote{For simplicity, we use the notation $\mathbf{n}\equiv\mathbf{n}(\mathbf{s})$ (and thus $n_1\equiv n_1(s_1,\dots,s_N),\dots,  n_q\equiv n_q(s_1,\dots,s_N)$) hence the dependence on $\mathbf{s}$ will be assumed and dropped henceforth. We will also use the notations for the sums $\sum_{\mathbf{s}}\equiv \sum_{s_1}\dots \sum_{s_N}$ and $\sum_{\mathbf{n}}\equiv \sum_{n_1\ge 0}\dots \sum_{n_q\ge 0}\delta_{\sum_s n_s,N}$, where $\delta_{i,j}$ is the Kronecker delta.}. Here, we focus on protocols that  rely on the change of model parameters ($\Bell$ here). The cyclic operation of the engine is as follows (see Fig.~\ref{FigCycle}):
 \begin{enumerate}[A)]
 \item Initially, particles are distributed uniformly in the box according to the equilibrium distribution. In terms of the variables $\mathbf{s}$, the equilibrium state corresponds to the distribution
 \begin{equation}
\label{model}
P_{\rm eq}(\mathbf{s}|\Bell)= \prod_{i=1}^N \ell_{s_i} = \prod_{s=1}^q\ell_s^{n_s}.
\end{equation} 
In other words, in the equilibrium state each $s_i$ is drawn independently from the same distribution, with $\ell_s$ being the probability that $s_i=s$.
\item  Next, the system is driven quasistatically into an arbitrary  non-equilibrium distribution $P_{\rm neq}(\mathbf{s})$.
\item In the next step, the demon inserts $q-1$ rigid walls at positions $y_1,\ldots,y_{q-1}$, separating the box in $q$ partitions. 
When {the} walls are inserted, the positions of the particles   $\mathbf{s}$ remain constrained, i.e.,  frozen, for the rest of the cycle.
The demon measures the number  of particles in each partition $\mathbf{n}$. Note that the walls are introduced in the {\em same} positions in each cycle.
\item The demon applies a feedback protocol, based on the measurement $\mathbf{n}$, extracting work from the reversible expansion of the different partitions from their initial sizes $\Bell$ to new sizes $\Bell'$.
\item The cycle is completed by removing the walls from $\Bell'$  and resetting the system to the initial equilibrium state (A). Once the partitions are removed, the constraints are dropped and $s_i$ ($i=1,\dots,q$) become again dynamical variables.
\end{enumerate}

We shall denote by $\langle f(\mathbf{s})\rangle_{\rm neq}=\sum_{\mathbf{s}} P_{\rm neq}(\mathbf{s})f(\mathbf{s})$, averages over a non-equilibrium distribution $P_{\rm neq}$, and similarly $\langle f(\mathbf{s})\rangle_{\rm eq}=\sum_{\mathbf{s}} P_{\rm eq}(\mathbf{s})f(\mathbf{s})$. We denote by $W_{\rm ext}(\mathbf{n})$ the stochastic work extracted in each operation of the cycle\footnote{We take here the thermodynamic convention by which $W>0$ then work is done on the system, and $W<0$ when it is extracted from the system. We thus define $W_{\rm ext}=-W$.} and $\langle W_{\rm ext}\rangle_{\rm neq}$ its average over many cycles.

Previous work~\cite{song2019optimal} discussed work extraction from equilibrium states of the form~(\ref{model}), for different partition sizes $\Bell$. This corresponds to the cycle A$\to$C$\to$D$\to$E$\to$A (see Fig.~\ref{FigCycle}), which does not include the non-equilibrium driving A$\to $B. 
Ref.~\cite{song2019optimal} showed that{:} (i) $\langle W_{\rm ext}\rangle_{\rm eq}$ depends
on the initial partition sizes $\Bell$ and 
(ii) the optimal choice of $\Bell$ allows to extract an amount of work $-$ from an initial equilibrium state $-$ that saturates at $\sim 0.8371(q-1)$ for $N$ large. Apart from the case $N=1, q=2$, this protocol does not allow to saturate the 
second law inequality with feedback control~\cite{SagawaUeda2010}
$\langle W_{\rm ext}\rangle_{\rm eq} \le I(\mathbf{x};\mathbf{n})$. It was shown that it is possible to saturate the inequality with different protocols if the particles   interact weakly~\cite{Horowitz2011}.

\subsection{Work extraction}
\label{sec:we}

We now discuss the following optimization problem: what is the maximum work that can be extracted   from an average over many cycles of an $N$-particle Szil\'ard engine, and for which   {\em non-equilibrium} distribution the maximum average work extraction is achieved? 

In our analysis, we will consider the work done by insertion (B$\to$C) and removal (D$\to$E) of the walls as negligible~\cite{parrondo2001szilard}. In addition 
the work $W_{\rm E\to A}$ needed to restore the initial state is also zero, since once the walls are removed the partitions have no mechanical effect on the underlying gas.
Thus the total work done 
has two contributions: $W_{\rm A\to B}$ corresponding to the preparation of the non-equilibrium state, and $W_{\rm C\to D}$ corresponding to the expansion of the walls $\Bell\to\Bell'$.

The preparation of the non-equilibrium 
state requires an amount of work that  is
at least the free energy difference between the non-equilibrium distribution $P_{\rm neq}$ and the equilibrium state $P_{\rm eq}$, given by the Kullback-Leibler (KL) divergence~\cite{esposito2011second}\footnote{This equation assumes that the particles 
in each partitions are in an equilibrium state, i.e.\NM{,} 
$P(\hat x|\mathbf{s})=A(\mathbf{s})\prod_{s=1}^q \delta_{s_i,s(x_i)}$, with $A(\hat
s)$ {being} a normalisation constant. 
}:
\begin{equation}
\label{DF}
\langle W_{\text{A}\to\text{B}}\rangle\geq\Delta F_{\text{A}\to\text{B}}=D\left[P_{\rm neq}(\mathbf{s})||P_{\rm eq}(\mathbf{s}|\Bell)\right]=\sum_{\mathbf{s}}P_{\rm neq}(\mathbf{s})\ln\frac{P_{\rm neq}(\mathbf{s})}{P_{\rm eq}(\mathbf{s}|\Bell)}.
\end{equation}

To calculate the work done in the expansion C$\to$D, we assume the law of ideal gases and unit volume, hence the initial pressure inside the box is $\mathsf{P}_{\rm in}=N$. The insertion of the walls by the demon  puts the system in a state where the gas is in local equilibrium in each partition 
with a different pressure, $\mathsf{P}_s=n_s/\ell_s$ ($s=1,2,\ldots,q$). 
The forces exerted on the internal walls can be exploited to extract work, 
using the knowledge (i.e. measurement) of $\mathbf{n}$. These forces will
vanish when the walls adjust such  that the
pressures in all partitions equalise and revert to the initial
values, i.e.{,}
$\mathsf{P}'_s=n_s/\ell_s'=N$. This implies that the new
partition sizes are given by $\hat\ell_s=n_s/N$. 
The work $-W_{\rm C\to D}(\mathbf{n})$ extracted from the expansion of the gas can be written in two ways~\cite{song2019optimal}:
\begin{eqnarray}
-W_{\rm C\to D}(\mathbf{n}) & = & \sum_{s=1}^q\int_{\ell_s}^{n_s/N}\text{d}V\frac{n_s}{V}=\sum_{s=1}^q n_s\ln\frac{n_s}{N\ell_s} \\
 & = & \ln\frac{P_{\rm eq}(\mathbf{s}|\hat{\Bell})}{P_{\rm eq}(\mathbf{s}|\Bell)}= \max_{\Bell'}\;\ln\frac{P_{\rm eq}(\mathbf{s}|\Bell')}{P_{\rm eq}(\mathbf{s}|\Bell)}.\label{maxlik}
\end{eqnarray}
Eq. (\ref{maxlik}) implies that 
optimal  work extraction is equivalent to adjusting the parameters $\hat\Bell$ so as to maximise the likelihood of $\mathbf{s}$ 
under model Eq. (\ref{model}). The solution
\begin{equation}
 \hat\ell_s(\mathbf{s})\equiv{\rm arg}\max_{\Bell'}\ln P_{\rm eq}(\mathbf{s}|\Bell') = \frac{n_s}{N},
\end{equation}
coincides with the maximum likelihood estimators (MLE) of the parameters $\ell_s$~\footnote{As done for $\mathbf{n}$, we will drop in the following for simplicity the dependency of $\hat\Bell(\mathbf{s})$  on $\mathbf{s}$ and simply write $\hat\Bell$.}.
After this re-adjustment phase, the 
partitions are removed and the system's original
equilibrium state {is restored}.

A cycle of the information engine includes the preparation of the 
non-equilibrium state and the work extraction phase.  
The total {amount of}
work that can be extracted from a cycle is at most
\begin{eqnarray}
\langle W_{\rm ext}\rangle_{\rm neq}&\leq &-\Delta F_{\text{A}\to\text{B}}-\langle W_{\rm C\to D}\rangle_{\rm neq}\\
&=& -\sum_{\mathbf{s}}P_{\rm neq}(\mathbf{s})\ln\frac{P_{\rm neq}(\mathbf{s})}{P_{\rm eq}(\mathbf{s}|\Bell)}+\sum_{\mathbf{s}}P_{\rm neq}(\mathbf{s})\ln\frac{P_{\rm eq}(\mathbf{s}|\hat\Bell)}{P_{\rm eq}(\mathbf{s}|\Bell)}\\
&=&-\sum_{\mathbf{s}}P_{\rm neq}(\mathbf{s})\ln\frac{P_{\rm neq}(\mathbf{s})}{P_{\rm eq}(\mathbf{s}|\hat\Bell)}. \label{eq:10} 
\end{eqnarray} 
We remark here that $P_{\rm eq}(\mathbf{s}|\hat\Bell)$ is not normalized because $\hat\Bell$ depends on $\mathbf{s}$, through $\mathbf{n}$. The corresponding normalized distribution reads
\begin{equation}
\label{nmlmn}
 P^*(\mathbf{s})\equiv \frac{P_{\rm eq}(\mathbf{s}|\hat\Bell(\mathbf{s}))}{\sum_{\mathbf{s}'}P_{\rm eq}(\mathbf{s}'|\hat\Bell(\mathbf{s}'))}=e^{-\mathcal{R}^{\star}} \prod_{s=1}^q \left(\frac{n_s}{N}\right)^{n_s}.
 \end{equation}
The parameter  $\mathcal{R}^{\star}$ in the normalization constant $e^{-\mathcal{R}^{\star}}$ is given by
\begin{equation}
\label{StocCompl0}
\mathcal{R}^{\star}  =  \ln \left[ N!\sum_{\mathbf{n}}\prod_{s=1}^q \frac{(n_s/N)^{n_s}}{n_s!}\right],
\end{equation}
where the sum in~(\ref{StocCompl0}) is restricted to all vectors of  positive integers $\mathbf{n}$ such that $\sum_{s=1}^q n_s=N$.
On substituing Eqs.~(\ref{nmlmn}) and~(\ref{StocCompl0}) in Eq.~(\ref{eq:10}), we obtain
\begin{equation}
\langle W_{\rm ext}\rangle_{\rm neq} \leq \mathcal{R}^{\star} - D[ P_{\rm neq}(\mathbf{s})||P^{\star}(\mathbf{s})] \leq  \mathcal{R}^{\star} ,
\end{equation}
where the first inequality saturates when the process A$\to$B is quasistatic. In this case, the maximum average work, $\langle W_{\rm ext}\rangle_{\rm neq}=\mathcal{R}^{\star}$, is attained when $D[ P_{\rm neq}(\mathbf{s})||P^{\star}(\mathbf{s})]=0$, i.e.  $P_{\rm neq}(\mathbf{s})=P^{\star}(\mathbf{s})$, and thus when  the non-equilbrium distribution is given by the distribution~(\ref{nmlmn}). Note that 
neither $P^*$ nor $\mathcal{R}^{\star}$ depend on the initial state $\Bell$ of the engine.

We observe that the information engine does not measure
the whole sample $\mathbf{s}$, but only the counts $\mathbf{n}$, which are the {\em sufficient statistics} of the equilibrium model  (\ref{model}). 
The constant $\mathcal{R}^*$ can be estimated asymptotically when $N\gg q$ (see e.g. \cite{beretta2018stochastic}), and
\begin{equation}
\mathcal{R}^{\star}\simeq \frac{q-1}{2}\ln N+\frac{q}{2}\ln\pi -\ln\Gamma(q/2).
\label{UC_SZ}
\end{equation}
Therefore, the work extracted in the cycle grows {with $N$} as $[(q-1)/2]\ln N$. This contrasts with the case where work is extracted from the equilibrium state, where the work depends on $\Bell$ and the maximal work is bounded above by a constant, when $N\to\infty$~\cite{song2019optimal}. 

\MM{For thermodynamic processes with feedback control, the second law of thermodynamics~\cite{sagawa2012thermodynamics} sets a limit to the amount of  work that can be extracted, which cannot exceed the entropy of the measured quantity. In the present case, this bound reads $\langle W_{\rm ext}\rangle \le k_{\rm B} T\,H[\bn]$, where 
\begin{equation}
H[\bn]=-\sum_{\bn}P^*(\bn)\ln P^*(\bn),
\end{equation}
is the entropy of the measurement $\bn$, and
\begin{equation}
P^*(\bn)=
e^{-\mathcal{R}^{\star}} N!\prod_{s=1}^q \frac{\left({n_s}/{N}\right)^{n_s}}{n_s!}.
\end{equation}
It is interesting to see how close the protocol discussed above comes to this limit. 
In the limit $N\to\infty$, we find (see Appendix \ref{appH})
\begin{equation}
\label{Hnasymp}
H[\bn]\simeq (q-1)\ln N+\frac q 2 \ln \pi -\ln \Gamma\left(\frac q 2 \right)+q[\psi(q/2)-\psi(1/2)],\qquad N\gg q,
\end{equation}
where $\psi(z)=\Gamma'(z)/\Gamma(z)$ is the digamma function. Taking Eq. \eqref{UC_SZ} into account, we find that
\begin{equation}
\label{eqeta}
\lim_{N\to \infty} \frac{\mathcal{R}^*}{H[\bn]}=\frac 1 2.
\end{equation}
This result is reminiscent of those derived in Refs.~\cite{hartich2016sensory,matsumoto2018role} on sensory capacity of bipartite systems. We observe that Eq. \eqref{eqeta} has a simple interpretation in information theoretic terms, in the present setting: When $\bn$ is measured, its entropy  is reduced from $H[\bn]\simeq (q-1)\ln N$ nats to zero. After the walls are removed, the 
uncertainty on the occupation numbers $\bn$ increases. Indeed in state E, $n_s$ are known to a relative precision $\delta n_s/n_s\sim 1/\sqrt{N}$, which requires $\frac{q-1}{2} \ln N$ nats\footnote{Remember that one of the occupation numbers is fixed by the constraint $\sum_s n_s=N$}. Hence half of the information harvested in the measurement is lost when the walls are released (D$\to$E).}
 
\begin{figure}[h!]
\centering
\includegraphics[width=0.9\linewidth]{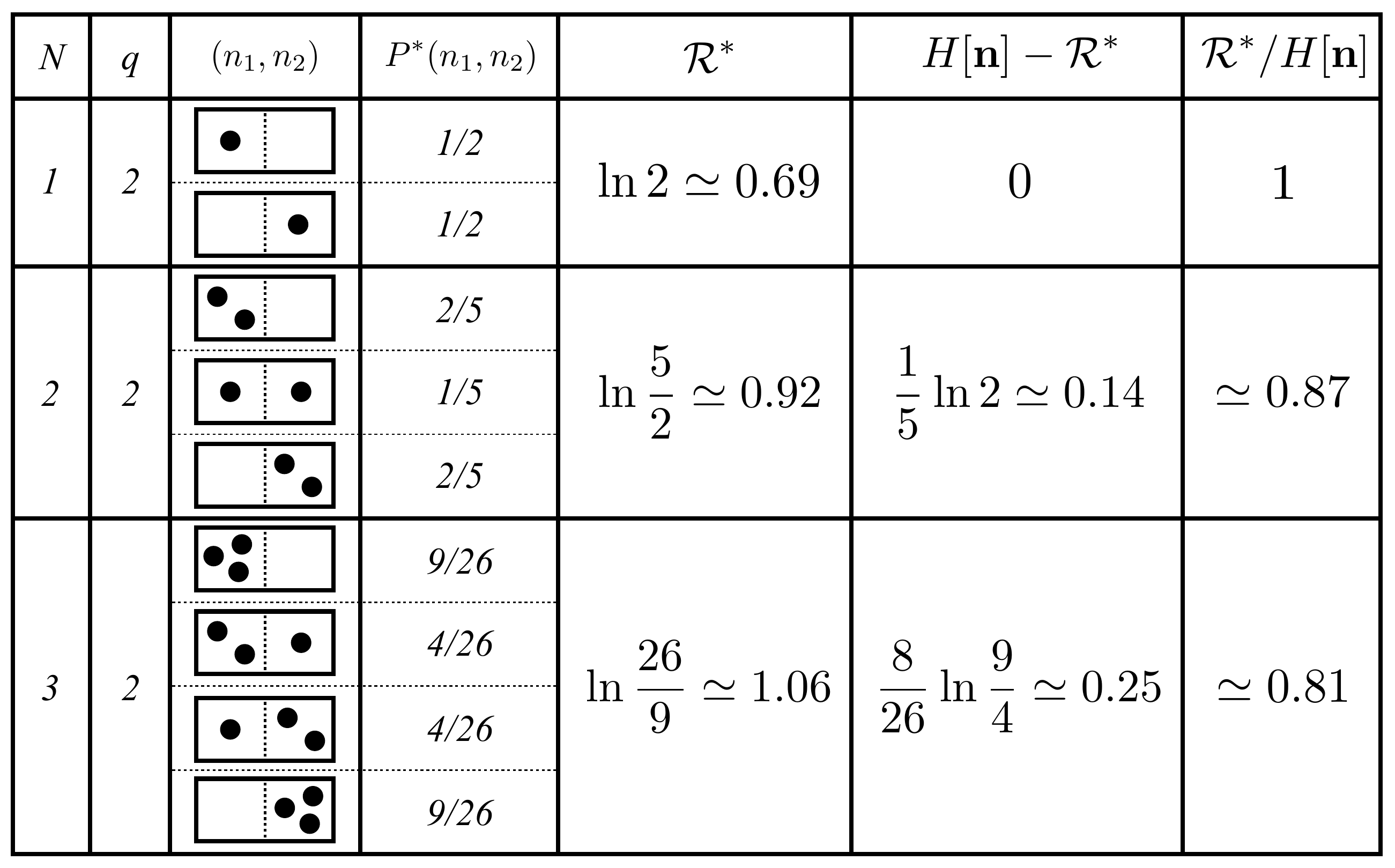}
\caption{\small Example values for the optimal non-equilibrium distribution ($P^{\star}$, Eq.~(\ref{nmlmn})) and its corresponding maximum average work in $k_{\rm B}T$ units ($\mathcal{R}^{\star}$, Eq.~(\ref{StocCompl0})) for $N=1,2,3$ particle Szil\'ard engines with $q=2$ partitions of equal size. The fourth column reports the probability $P^*(n_1,n_2)$ of finding $n_1$ and $n_2$ particles on the left and on the right partitions, respectively. }
\label{fig:szexamples}
\end{figure}

 We exemplify our result in Fig.~\ref{fig:szexamples} by providing the values of the optimal non-equilibrium distribution $P^*$ $-$Eq.~(\ref{nmlmn})$-$ its corresponding average extracted work $k_{\rm B}T \mathcal{R}^{\star}$   $-$Eq.~(\ref{StocCompl0})$-$ 
 \MM{and the entropy $H[\bn]$}, 
 in Szil\'ard engines with $q=2$ partitions of equal size $\ell_1=\ell_2=1/2$ for three different number of particles, $N=1,2,3$. The $N=1$ particle Szil\'ard engine is the only example for which $P^{\star} = P_{\rm eq}$ and thus $\mathcal{R}^*=H[\bn]=\ln 2$, as expected. The maximum extractable work is $k_{\rm B}T\ln (5/2)$ for $N=2$ and $k_{\rm B}T\ln (26/9)$ for $N=3$. This is 32\% and 81\% larger than the work that can be extracted from an initial equilibrium state for $N=2$ and $N=3$, respectively~\cite{song2019optimal}, \MM{and it is 87\% and 81\% of the entropy of the measurement $H[\bn]$, respectively.}
 Notably, for both $N=2$ and $N=3$ Szil\'ard engines, we find that the optimal $P^{\star}(\bs)$ is such that states where particles are distributed asymmetrically, i.e. with particles clustered in one of the boxes, are considerably more likely than under the equilibrium distribution\footnote{\MM{We observe that, when all the particles are in the same partition, $P_{\rm eq}(\bn|\hat\Bell)$ attains its maximal value, i.e. $\max_{\bn}P_{\rm eq}(\bn|\hat\Bell)=1$. This implies that the maximal work that can be extracted equals the min-entropy, i.e. $\cR^*=-\ln \max_{\bn} P^*(\bn)$. The min-entropy, which is the R\'enyi entropy of infinite order, appears in optimization problems of "one-shot" thermodynamics~\cite{halpern2015introducing}. Fig.~\ref{fig:szexamples} illustrates this relation for $N=1,2,3$ and $q=2$.} \RR{Note also that the maximal work extracted in the transition from C to D is when all particles are found in the smallest partition. Yet the work needed to prepare the system in that state is exactly equal to the work that can be extracted, i.e. it is $-N \log(\min_s\ell_s)$ in both cases. So the net work extracted over a cycle would be zero. This strategy is inefficient both from the thermodynamic point of view and from the one of coding. Indeed, this would correspond to a degenerate (zero entropy) source, where the probabilities are $1$, for one of the partitions, and $0$ for all the others.}}. This result suggests a well-defined procedure to achieve optimal work in Szil\'ard engines by means of attractive interaction potentials between the  particles. 

\subsection{Relation to the Minimum Description Length (MDL) principle}

Equations~(\ref{nmlmn}) and~(\ref{StocCompl0}) are 
intimately related to the solution of an apparently very different problem,
{ the problem} of optimal {universal} coding, which sets the stage to the MDL principle. 
We limit our discussion to a brief account and refer to Ref.\
\cite{grunwald2007minimum} for a {more} detailed discussion. 

Let $\mathbf{s}=(s_1,\dots,s_N)$ be a message generated by an
information source that we wish to compress as much as possible.
This entails {to} translating $\mathbf{s}$ in{to} a sequence of nats\footnote{For later convenience, 
we measure information in nats and take logarithms in the natural basis.}. A code is a rule that assigns to each 
$\mathbf{s}$ a sequence of nats {by a one--to--one mapping}. 
The {\em description length}\NM{,} $E(\mathbf{s})$\NM{,} is the number of nats of
the compressed sequence. The key to achieve efficient codes is to assign short codewords to 
frequent patterns. This has to be done satisfying Kraft-McMillan inequality
\begin{equation}
\label{Kraft}
\sum_{\mathbf{s}} e^{-E(\mathbf{s})}\le 1,
\end{equation}
which imposes a fundamental limit to uniquely decodable codes~\cite{cover2012elements}.


Let us consider messages $\mathbf{s}$ that are generated by drawing at random and independently each character $s_i$ 
from a probability distribution, $f({s_i}|\Btheta)$, that depends 
on a vector of $M\ge 1$ parameters $\Btheta=(\theta_1,\dots,\theta_M)$. 
If the parameters $\Btheta$ are known, the optimal description length that can be
achieved {(ignoring integer length constraints)} is~\cite{cover2012elements} 
\begin{equation}
E(\mathbf{s}|\Btheta) = - \ln P(\mathbf{s}|\Btheta)\equiv -\sum_{i=1}^N\ln f(s_i|\Btheta).
\label{energy}
\end{equation}
MDL deals with optimal coding in the case where $\Btheta$ are unknown. Then, 
for a given message $\mathbf{s}$, the optimal compression is achieved minimising the
description length over $\Btheta$, i.e. $E(\mathbf{s}|\hat\Btheta)=\min_{\Btheta} \{-\ln P(\mathbf{s}|\Btheta)\}$. 
A code with description length $E(\mathbf{s})$ entails to an excess code length for this sample, of
$\mathcal{R}(E,\mathbf{s})=E(\mathbf{s})-E(\mathbf{s}|\hat\Btheta)$ nats, 
which is called the {{\em redundancy}\footnote{\MM{$\cR$ is also called {\em regret} in the
literature of predictive learning theory, in the context of learning with expert advice.}}.
{The} MDL identifies optimal codes with those achieving the minimal redundancy, for the worst possible sample~\cite{grunwald2007minimum}
\begin{equation}
\mathcal{R}^{\star}=\min_{E}\max_{\mathbf{s}}\mathcal{R}(E,\mathbf{s}),
\end{equation}
where the minimum is taken over all $E(\mathbf{s})$ 
satisfying the Kraft-McMillan inequality~(\ref{Kraft}).
A code that solves this problem
is {associated with the length function,} $E^*(\mathbf{s})=-\ln 
P^*(\mathbf{s})${,} where
\begin{equation}
\label{NMLgeneral}
P^*(\mathbf{s})=e^{-\mathcal{R}^{\star}}P(\mathbf{s}|\hat\theta (\mathbf{s})),
\end{equation}
{which} is called the normalised maximum likelihood (NML) {probability
distribution} and 
\begin{eqnarray}
\label{UC_redundancy}
\mathcal{R}^{\star} & = & \ln\sum_{\mathbf{s}}P(\mathbf{s}|\hat\theta(\mathbf{s})) \\
 & \simeq & \frac{M}{2}\ln\frac{N}{2\pi}+\int \text{d}\Btheta \sqrt{{\rm det}\, {J}(\Btheta)}, \label{cexp}
\end{eqnarray}
is \NM{the \emph{minimax universal code redundancy}}~\cite{rissanen1996fisher}, \NM{that we shall call minimax redundancy for short, in what follows}. 
Equation~(\ref{cexp}) provides the large $N$ leading behaviour~\cite{rissanen1996fisher}, where $M$ is the number of parameters (i.e., the dimension of the vector $\Btheta$) and  ${J}(\Btheta)$ is the Fisher information matrix. The latter is a $M\times M$ matrix with elements 
\begin{equation}
\label{FIM}
J_{a,b}(\Btheta)=\sum_{\bs}P(\bs|\Btheta)\left[\frac{\partial}{\partial\theta_a}\ln   P(\mathbf{s}|\Btheta)\right]
\left[\frac{\partial}{\partial\theta_b}\ln   P(\mathbf{s}|\Btheta)\right].
\end{equation}
Notice
that, a sample $\mathbf{s}$ of {length} $N$ allows to estimate each parameter 
to a precision {of} ${\delta\theta_a}\sim 1/\sqrt{N}$, and hence it provides 
$\frac{1}{2}\ln N$ nats of information for each parameter, in agreement with
the first term of Eq.\ (\ref{cexp}). 
In a loose sense, $\mathcal{R}^{\star}$ quantifies the amount of information that the sequence $\mathbf{s}$ 
provides on the parameters $\Btheta$ of the model from which it has been drawn.\footnote{This statement can be made more precise
within a Bayesian framework, where Eq.\ (\ref{cexp}) is the asymptotic value 
of the mutual information $I(\mathbf{s};\theta)$ between a sample and the
parameters, under Jeffrey's prior.}

The NML provides a {\em universal code} because it achieves a compression per data point 
which is as good as the compression that would be achieved asymptotically as $N\to\infty$ with the optimal choice of $\Btheta$.  Indeed, as shown by Rissanen \cite{rissanen1984universal}, $\mathcal{R}^{\star}$ 
grows only as $\ln N$ (Eq. \ref{cexp}), so ${\mathcal{R}^{\star}}/N\to 0$ as $N\to \infty$~\footnote{Moreover, the convergence of ${\mathcal{R}^{\star}}/N$ to zero is at the fastest
possible rate when $\Btheta$ is unknown (in a very strong sense) \cite{rissanen1984universal}.}. Note that the minimax redundancy, ${\mathcal{R}^{\star}}$ in Eq.~(\ref{UC_redundancy}) is independent of the sample $\mathbf{s}$. Therefore it provides a measure of the complexity of the model $f$, which can be used in model selection. 

Eq.\ (\ref{model}) is a parametric model that depends on the parameters $\theta_s=\ell_s$, $s=1,\ldots,q-1$ with $M=q-1$ ($\ell_q=1-\ell_1-\ldots -\ell_{q-1}$ fixed by the normalisation constraint). Its \NM{minimax redundancy}~(\ref{UC_redundancy}) 
coincides with the maximal average  work extracted in a cycle of the $N$-particle Szil\'ard engine~(\ref{StocCompl0}), and the NML~(\ref{NMLgeneral}) coincides with the optimal non-equilibrium distribution $P^{\star}$ in Eq. (\ref{nmlmn}). This  is the main result of this paper; it connects information thermodynamics with the MDL principle.  We provide in Table~\ref{tab:1} a "dictionary" of  analogies between different  quantities in information thermodynamics and  coding  that stem from our theory.

In the rest of the paper, we shall see that the subtle connection between thermodynamics and coding  unveiled here  is a very 
general result that applies to a broad class of parametric models. 
As we will see, this is most easily seen if the original Szil\'ard engine is described in 
terms of a model of a system of non-interacting classical particles in  $q$ energy levels. 

\begin{table}[h!]
\centering
\begin{tabular}{|l|l|}
\hline 
 {\bf information engine} & {\bf coding (MDL)} \\
\hline 
\hline 
 microscopic state $\mathbf{s}$ & message $\mathbf{s}$ \\
\hline 
 particle coordinate $s_i$ & character $s_i$\\
\hline 
measurement $\mathbf{n}$ ($\mathbf{\Phi}$) & sufficient statistics $\mathbf{n}$ ($\mathbf{\Phi}$) \\
\hline 
protocol parameters $\Bell$ & model's parameters $\Btheta$\\
\hline 
equilibrium state $P_{\rm eq}(\mathbf{s}|\Bell)$ & parametric model $P(\mathbf{s}|\Btheta)=e^{-E(\mathbf{s}|\Btheta)}$ \\
\hline 
non-equilibrium state $P_{\rm neq}(\mathbf{s})$ & (exponentiated) coding function $e^{-E(\mathbf{s})}$\\
\hline 
optimal non-equilibrium state $P^*(\mathbf{s})$ & optimal coding distribution $P^*(\mathbf{s})$ (NML)\\
\hline 
maximal work extracted $\langle W_{\rm ext}\rangle$ & minimax universal code redundancy $\mathcal{R}^*$ \\
\hline 
energy levels $\epsilon_s=-\ln \ell_s$ & description length $\epsilon_s$ for character $s$\\
\hline 
\end{tabular}
\caption{Correspondence  between  different concepts in information engines and in MDL coding.}
\label{tab:1}
\end{table}

\newpage
\section{Physical implementation and further extensions}
\label{sec:iii}

\subsection{A physical realization of  the $N$-particle Szil\'ard engine}

We now discuss, in the light of stochastic thermodynamics, a physical analog of a Szil\'ard engine which enlightens extensions of our theory to more general scenarios.  The model consists of  $N$ classical non-interacting particles in $q$ energy levels $\Bepsilon=(\epsilon_1,\dots,\epsilon_q)$ immersed in a thermal bath with $k_{\rm B}T=1$.  We let $s_i$ denote the energy level occupied by the $i^{\rm th}$ particle and $n_s$ be occupation number of the $s^{\rm th}$ energy level. Each level can be occupied by a maximum of $N$ particles, with the constraint $\sum_{s=1}^{q} n_s=N$, i.e. conservation of the total number of particles. 
The equilibrium probability of a configuration $\mathbf{s}=(s_1,\dots,s_N)$ is 
\begin{equation}
    P_{\rm eq}(\mathbf{s}|\Bepsilon)=\frac{e^{-\mathcal{H}(\mathbf{s})}}{Z(\Bepsilon)}=e^{-[\mathcal{H}(\mathbf{s})-\mathcal{F}_{\rm eq}(\Bepsilon)]},
\label{model_e}
\end{equation}
where $\mathcal{H}(\mathbf{s})=\sum_{i=1}^N\epsilon_{s_i}=\sum_{s=1}^{q} n_s\epsilon_s$ is the total energy (Hamiltonian) of the particles,  $Z(\Bepsilon) = \left(\sum_{s=1}^q e^{-\epsilon_s}\right)^N$ the partition function, and $\mathcal{F}_{\rm eq}(\Bepsilon)=-\ln Z(\Bepsilon)$ the equilibrium free energy. Setting 
\begin{equation}
\ell_s=\frac{e^{-\epsilon_s}}{\sum_{s'=1}^q e^{-\epsilon_{s'}}}
\end{equation} for all $s=1,\dots,q$, Eq. (\ref{model_e}) becomes identical to Eq.~(\ref{model}), where $\ell_s$ is the probability that a particle is found in partition $s$ of the Szil\'ard engine. Hence  the system of particles described by Eq. (\ref{model_e}) is formally equivalent to a 
Szil\'ard engine with a box of size $L=\sum_{s=1}^q e^{-\epsilon_s}$ in the $x$-direction and partitions of size $e^{-\epsilon_s}$. 
Note that $L=1$ can be enforced by an appropriate shift of all energy levels and that would imply $\mathcal{F}_{\rm eq}(\Bepsilon)=0$.

\begin{figure}[h!]
\centering
\includegraphics[width=0.7\linewidth]{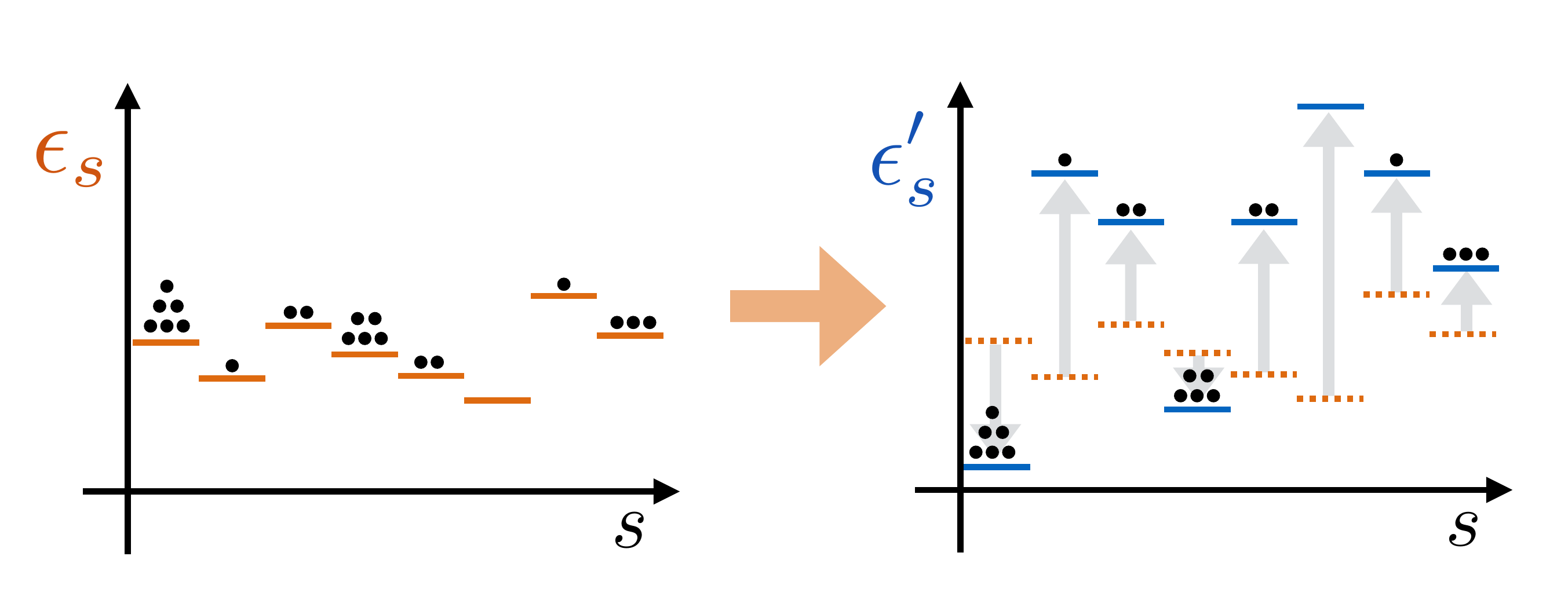}
\caption{\small Re{-}adjustment of energy levels during 
the work extraction transformation C$\to$D. The occupation of energy
levels\NM{,} $\mathbf{n}$\NM{,} stays the same during this transformation.}
\label{fig:workextr}
\end{figure}

Within this analogy,  the non-equilibrium driving in the Szil\'ard engine A$\to$B  requires an amount of work of at least, see Eq. (\ref{DF}), 
\begin{equation}
\langle W_{\rm A\to B}\rangle\geq\Delta F_{\rm A\to B}=\left\langle\ln P_{\rm  neq}(\mathbf{s})\right\rangle_{\rm neq}+\left\langle \mathcal{H}(\mathbf{s})\right\rangle_{\rm neq}-\mathcal{F}_{\rm eq}(\Bepsilon).
\end{equation}
The work extraction phase C$\to$D  corresponds to adjusting the energy levels  $\Bepsilon\to\Bepsilon'$
 without changing the population $\mathbf{n}$ of the levels (see Fig.\ \ref{fig:workextr}). 
Since particles do not jump between different levels, no heat is released in this transformation\footnote{Note that if the energy levels
$\{\epsilon_s'\}$ are such that $\sum_s e^{-\epsilon_s'}\neq 1$, this process also involves 
an expansion of the Szil\'ard box as a whole followed by a 
compression (or {\em vice versa}). This global expansion/compression does not occur in the optimal protocol.}, so the work equals the change in internal energy $W_{\rm C\to D}(\mathbf{s}) = \mathcal{H}'(\mathbf{s})-\mathcal{H}(\mathbf{n})$, where $\mathcal{H}'(\mathbf{s})=\sum_{i=1}^N\epsilon'_{s_i}=\sum_{s=1}^{q} n_s\epsilon'_s$. 

When the partitions are removed D$\to$E the occupation numbers $\mathbf{n}$ equilibrate to the new  energy levels without any work expenditure\footnote{When partitions are removed ($D\to E$), an amount  $Q(\mathbf{s})=\sum_s [\langle  n_s\rangle_{\rm eq'}- n_s ]\epsilon_s'$ of heat is released to the thermal bath, where the average $\langle\ldots\rangle_{\rm eq'}$ is taken on the equilibrium distribution with energies $\Bepsilon'$.} ($W_{\rm D\to E}=0$). The cycle is completed when the system is driven back to the initial equilibrium E$\to$A, which requires a work\footnote{We shall see that $W_{\rm E\to A}=0$ when the engine operates under the optimal protocol.} $W_{\rm E\to A}=\mathcal{F}_{\rm eq}(\Bepsilon)-\mathcal{F}_{\rm eq}(\Bepsilon')$.

 Taken all together, the work extracted in a cycle satisfies the inequality
\begin{eqnarray}
    W_{\rm ext}(\mathbf{n},\Bepsilon')& \le &-\Delta F_{\rm A\to B}-W_{\rm C\to D}(\mathbf{s})-W_{\rm E\to A}\\
    & = & -\left\langle\ln P_{\rm  neq}(\mathbf{s})\right\rangle_{\rm neq}-\left\langle \mathcal{H}(\mathbf{s})\right\rangle_{\rm neq}+\mathcal{H}(\mathbf{s})-\mathcal{H}'(\mathbf{s})+\mathcal{F}_{\rm eq}(\Bepsilon')\\
    & = & -\left\langle\ln P_{\rm  neq}(\mathbf{s})\right\rangle_{\rm neq}+\mathcal{H}(\mathbf{s})-\left\langle \mathcal{H}(\mathbf{s})\right\rangle_{\rm neq}+\ln P_{\rm eq}(\mathbf{s}|\Bepsilon')
    \label{workee}
\end{eqnarray}
where we have used the definition~(\ref{model_e}) in the last step.
The only term that depends on the new energy levels $\Bepsilon'$ in Eq. \eqref{workee} is the last one. Therefore the lower bound to the work extracted is maximal for 
\begin{equation}
    {\hat\Bepsilon}(\mathbf{s}) = 
    {\rm arg}\max_{\Bepsilon'}\left[
    \ln P_{\rm eq}(\mathbf{s}| \Bepsilon')\right].
    \label{MLE}
\end{equation}
Note that work optimization corresponds to maximizing the likelihood, i.e. setting $\epsilon_s'=\hat\epsilon_s(\mathbf{s})=-\ln (n_s/N)$ which are the MLE of the distribution \eqref{model_e} (see Eq.~\ref{maxlik}).
Also note that the optimal energy levels $\hat\Bepsilon$ depend on $\bs$ through the measurement $\mathbf{n}$. 

Taking the average of Eq. \eqref{workee} with $\Bepsilon'=\hat\Bepsilon$ over $P_{\rm neq}$, 
we find the following upper bound for the work extracted averaged over many cycles (see Eq.~\ref{eq:10}),
\begin{eqnarray}
\langle W_{\rm ext}\rangle &=& - \left\langle\ln \frac{P_{\rm neq}(\mathbf{s})}{P_{\rm eq}(\mathbf{s}|\hat\Bepsilon) }\right\rangle_{\rm neq}  \nonumber\\
&=& \mathcal{R}^{\star} - D[ P_{\rm neq}(\mathbf{s})||P^{\star}(\mathbf{s})] \leq  \mathcal{R}^{\star}, \label{eq:25}
\end{eqnarray}
where again 
\begin{equation}
   P^*(\mathbf{s})=e^{-\mathcal{R}^{\star}}P_{\rm eq}(\mathbf{s}|\hat\Bepsilon(\mathbf{s}))  ,  \label{eq:27}
\end{equation}
is the NML and 
\begin{equation}
    \mathcal{R}^{\star}=\ln\left[\sum_{\mathbf{s}}P_{\rm eq}(\mathbf{s}|\hat\Bepsilon(\mathbf{s}))\right].
\end{equation}
The first inequality in~(\ref{eq:25})  is saturated in the quasistatic limit whereas the second inequality in~(\ref{eq:25}) is saturated when the non-equilibrium state is drawn from the NML {distribution}  \eqref{eq:27}.

\subsection{Optimal work extraction measuring sufficient statistics: theory}
\label{sec:suff}

We now ask the question: how do these results generalise when the measured quantities $\BPhi$ are different from $\mathbf{n}$?
In order to address this question, we consider a model where the energy levels 
\begin{equation}
\label{elevelgen}
\epsilon_s(\Btheta)=\epsilon_s^0-\Btheta\cdot\Bphi(s) =
\epsilon_s^0-\sum_{m=1}^M \theta_m\phi_m(s),
\end{equation}
depend on the parameters $\Btheta=(\theta_1,\ldots,\theta_M)$ through the $M$ functions $\Bphi(s)=(\phi_1(s),\ldots,\phi_M(s))$. 
Here the dot denotes the scalar product and $\epsilon_s^0$ specifies the energy of level $s$ when it is unoccupied ($n_s=0$).
For a microscopic state $\bs$, the Hamiltonian now takes the value 
\begin{equation}
    \mathcal{H}(\mathbf{s}|\Btheta)= \mathcal{H}_0(\mathbf{s})-\Btheta\cdot\BPhi(\bs),
    \label{Htheta}
\end{equation}
with 
\begin{equation}
\mathcal{H}^0(\mathbf{s})=\sum_{i=1}^N \epsilon_{s_i}^0=\sum_{s=1}^q  n_s\epsilon_s^0, 
\end{equation}
and the vector
\begin{equation}
\qquad \BPhi(\bs)=\left(\sum_{s=1}^q\Bphi_1(s)n_s,\dots,\sum_{s=1}^q\Bphi_M(s)n_s\right).
\end{equation}
The equilibrium distribution  becomes 
\begin{equation}
P_{\rm eq}(\mathbf{s}|\Btheta) = 
e^{- \mathcal{H}^0(\mathbf{s})+\Btheta\cdot\BPhi(\mathbf{s})+ \mathcal{F}_{\rm eq}(\Btheta) }, 
 \label{model1}
 \end{equation}
 with the corresponding equilibrium free energy given by
 \begin{equation}
 \mathcal{F}_{\rm eq}(\Btheta) =  -\ln\left[\sum_{\mathbf{s}}e^{- \mathcal{H}^0(\mathbf{s})+\Btheta\cdot\BPhi(\mathbf{s})}\right]. 
 \end{equation}
From the viewpoint of mathematical statistics, the parametric probability distribution~(\ref{model1}) belongs to an {\em exponential family}. Exponential families have special properties in that they exhibit explicitly the sufficient statistics, i.e., those functions of the data whose knowledge is sufficient to estimate the parameters from a sample. In the present context, notice that $P_{\rm eq}(\bs|\Btheta)=\prod_{i=1}^N f(s_i|\Btheta)$ is the probability of a sequence $\bs$ of $N$ independent draws $s_i$ from the distribution $f(s|\Btheta)=e^{-\epsilon_s^0+\Btheta\Bphi(s)-\varphi(\Btheta)}$, with $\varphi(\Btheta)=\mathcal{F}_{\rm eq}(\Btheta)/N$. For this distribution, $\BPhi(\bs)$ is a sufficient statistics, in the sense that it contains all necessary information about $\Btheta$~\cite{cover2012elements}, i.e. the mutual information between $\bs$ and the parameters $\Btheta$ equals that between $\Btheta$ and $\BPhi$:
\begin{equation}
I(\bs;\Btheta)=I(\Btheta;\BPhi).
\end{equation}
Exponential families are unique, in that the Pitman-Koopman-Darmois theorem \cite{koopman1936distributions} ensures that the set of sufficient statistics of a parametric model is independent of $N$ only if the model has the exponential form of Eq. (\ref{model1}). As we shall see, this is a crucial step in the work extraction cycle from measurements of sufficient statistics. 

An information engine based on the model \eqref{model1} performs a cycle analogous to that shown in Fig.~\ref{FigCycle}, with the sole exceptions that here the parameters of the model are $\Btheta$ instead of $\Bell$ and the  values of the statistics $\BPhi(\mathbf{s})$ are measured instead of $\mathbf{n}(\mathbf{s})$. Furthermore,  work is extracted by adjusting the conjugate parameters $\Btheta\to\Btheta'$. The work extracted in a cycle can be calculated in exactly the same way as in the discussion leading to Eq. (\ref{workee}). 
Note that, by Eqs. (\ref{Htheta}, \ref{model1}), $W_{\rm ext}$ depend on $\bs$ only through the measured quantity $\Phi$.
The work extracted is maximal for 
\begin{equation}
    {\hat\Btheta}(\mathbf{s}) = 
    {\rm arg}\max_{\Btheta'}\left[
    \ln P_{\rm eq}(\mathbf{s}| \Btheta')\right]= {\rm arg}\max_{\Btheta'}\left[\Btheta\BPhi(\bs)-\mathcal{F}_{\rm eq}(\Btheta)\right]
    \label{MLE2}
\end{equation}
which is again the MLE.
The crucial point is that the measured quantity $\BPhi$ should provide sufficient information to compute the values $\hat\Btheta$ of the optimal parameters in the work extraction phase. Therefore $\BPhi$ must be a sufficient statistics of model Eq. (\ref{model1}) and, by the Pitman-Koopman-Darmois theorem \cite{koopman1936distributions}, $P_{\rm eq}(\mathbf{s}|\Btheta)$ needs to have the exponential form of Eq. (\ref{model1}), for this to be possible for all $N$. Taking the expected value of the optimal extracted work, following the same steps of the previous section, we arrive at 
\begin{equation}
\langle W_{\rm ext}\rangle \le \mathcal{R}^{\star} - D[ P_{\rm neq}(\mathbf{s})||P^{\star}(\mathbf{s})],
\label{eq:Wextgen}
\end{equation}
where $P^*$ is the NML distribution \eqref{NMLgeneral} and $\mathcal{R}^*$ is the minimax redundancy of Eq. \eqref{UC_redundancy}. 
This shows that the maximal work that can be extracted from a cycle, on average, equals the minimax redundancy $\mathcal {R}^*$, and that the non-equilibrium state that achieves this bound is the NML, i.e $P_{\rm neq}=P^*$. The only change, with respect to the previous section, is that the re-adjustment of energy levels $\Bepsilon(\Btheta)$ is constrained by the functional form of Eq. (\ref{elevelgen}), to those that can be achieved by fine-tuning the parameters $\Btheta$, conjugate to the measured quantity.

Also in this case, this result is independent of the equilibrium parameters $\Btheta$ in which the engine is initialised and the optimal work extracted grows asymptotically as $\frac{M}{2}\ln N$ with the number of particles, for $N\gg 1$ (see Eq. \eqref{cexp}).

\MM{Let us now see how close the maximal extracted work approaches the fundamental limit imposed by the second law of thermodynamics, $\langle W_{\rm ext}\rangle \le k_B T H[\BPhi]$~\cite{sagawa2012thermodynamics}, where $H[\BPhi]$ is the entropy of  the measurement. In the limit of large $N\gg M$, to leading order, we find (see Appendix \ref{appH})
\begin{equation}
H[\BPhi]\simeq
M\ln N+\ln\int \! d\Btheta \sqrt{{\rm det}J(\Btheta)}+
\frac{\int d\Btheta \sqrt{{\rm det}J(\Btheta)} \ln \sqrt{{\rm det}J(\Btheta)}}{\int \! d\Btheta \sqrt{{\rm det}J(\Btheta)}}.
\end{equation}
This confirms that, as in Eq. (\ref{eqeta}), $\mathcal{R}^*/H[\BPhi]\to 1/2$ as $N\to\infty$. The intuition at the basis of this result is the same as in Section \ref{sec:we}: The total reduction in the uncertainty on $\BPhi$ brough about by the measurement is $H[\BPhi]\simeq M\ln N$.  A part of this is lost when the walls are removed, and the system relaxes to state E. In state E, the relative fluctuations of each component of $\BPhi$ are of order $1/\sqrt{N}$, hence the loss amounts to $\frac 1 2 \ln N$ nats per parameter. This result tallies with the findings of Ref. \cite{matsumoto2018role} in bipartite systems, that the measurement of sufficient statistics implies losses in thermodynamic efficiency.}

\subsection{Optimal work extraction measuring sufficient statistics: examples}
\label{sec:suff2}

\begin{figure}
\centering
\includegraphics[width=0.6\linewidth]{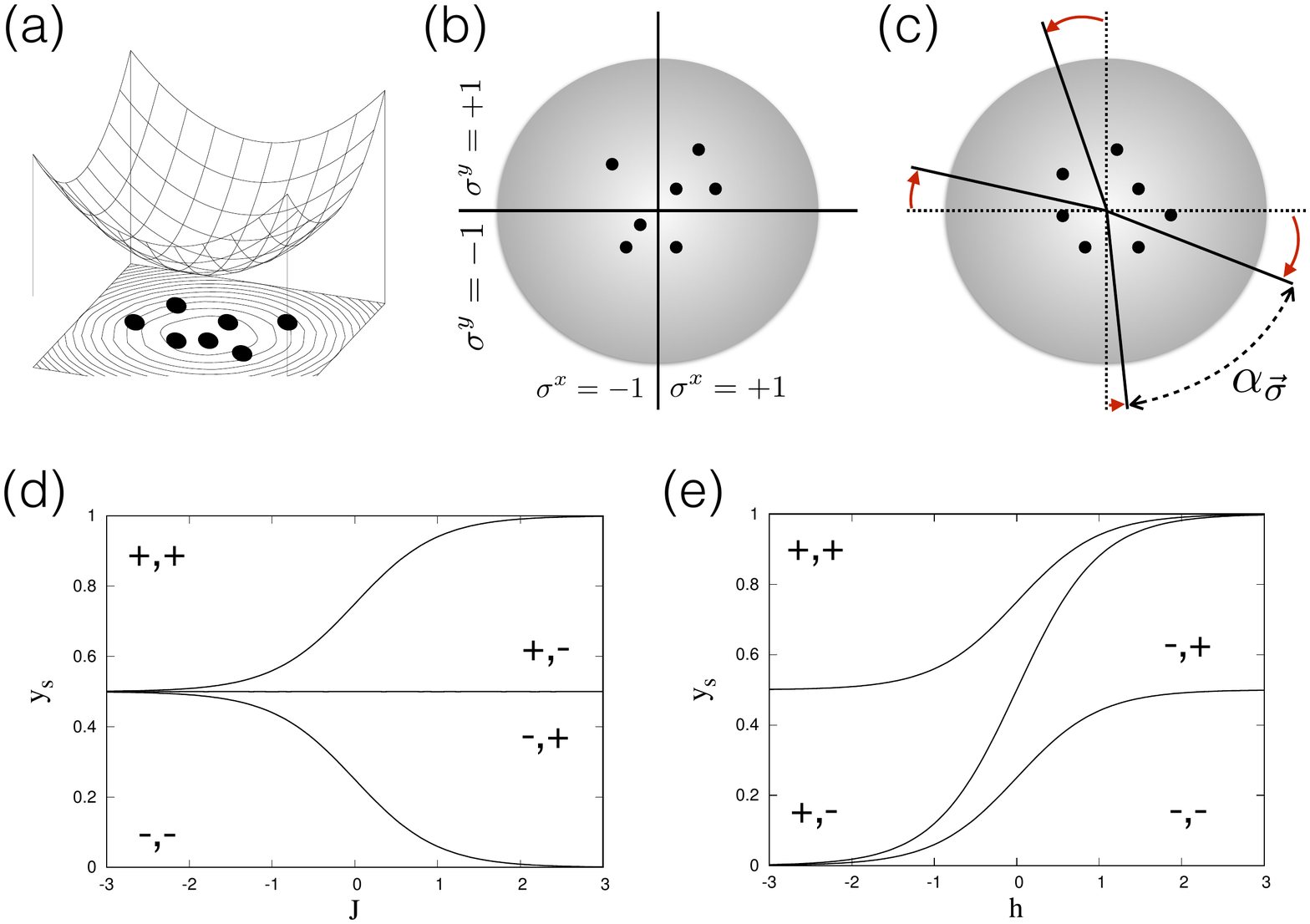}
\caption{\small \RR{Two examples of a Szil\'ard box with $q=4$ partitions. The system can be realised by an ideal classical gas of $N$ particles in 2D confined by a potential $V(r)$ that depends on the distance $r=\sqrt{x^2+y^2}$ from the origin (a). Walls can be inserted along the $x$ and $y$ axis (b) and allowed to rotate (c) to extract work. For the model in Eq. (\ref{spinJ}), the sizes of the partitions are shown in (d), as a function of the coupling constant $J$. Lines correspond to the angular coordinates $y_s$ ($s=++,+-,-+,--$) of the different walls for different values of the parameters. Panel (e) shows the analogous figure for the model in Eq. (\ref{spinh}).}
}
\label{figExample}
\end{figure}

\RR{Let us consider the case $q=4$ and  let us label the state of the $i$-th particle with a spin representation $s_i=\vec\sigma_i=(\sigma^{x}_i,\sigma^{y}_i)$, with $\sigma^{x}_i=\pm 1$ and $\sigma^{y}_i=\pm 1$ ($i=1,\dots,N$) . We let $\Bsigma=(\vec\sigma_1,\ldots,\vec\sigma_N)$ denote the state of the $N$ particle system in this representation. 
As shown in Fig. \ref{fig:szexamples}, this can be physically realised with an ideal gas of $N$ particles in two dimensions, confined by 
a potential $V(r)$, that increases with the distance $r=\sqrt{x^2+y^2}$ from the origin (Fig. \ref{fig:szexamples}(a)). The value of the spins identify the quadrant to which each article belongs. Work extraction proceeds by inserting walls along the two main axis (Fig. \ref{fig:szexamples}(b)), and by rotating them around the origin, as shown in Fig. \ref{fig:szexamples}(c). For this system, partition sizes correspond to the angle $\alpha_{\vec{\sigma}}=2\pi\ell_{\vec{\sigma}}$ between the walls delimiting the $\vec{\sigma}$ partition, as discussed in Appendix \ref{appspin}. }

For the model
\begin{equation}
\label{spinJ}
P_{\rm eq}(\Bsigma|J)=\frac{e^{J\Phi(\Bsigma)}}{(4\cosh J)^N},\qquad \Phi(\Bsigma)=\sum_{i=1}^N\sigma^{x}_i\sigma^{y}_i,
\end{equation}
the  box sizes and the energy levels  for a given value of the coupling constant $J$ are given by
\[
\ell_{\vec{\sigma}}(J)=\frac{e^{J\sigma^{x}\sigma^{y}}}{4\cosh J},\qquad \epsilon_{\vec{\sigma}}(J)=-J\sigma^x \sigma^y ,
\]
respectively.  As shown in Fig. \ref{figExample}(d),  ${\ell}_{\vec{\sigma}}(J)$ and $\epsilon_{\vec{\sigma}}(J)$ cannot  be adjusted independently 
In particular, under this model, the size of the partitions $[-,+]$ and $[+,-]$ remain equal in the work extraction protocol. This can be realised with the two wall on one of the axis staying fixed and the other two walls rotating together. 
Optimal work extraction demands that the parameter $J$ be adjusted to the MLE value 
\begin{equation}
\hat J(\Bsigma)={\rm arg}\max_J\left[J\BPhi(\Bsigma)-\ln(4\cosh J)\right]=
{\rm arg}\tanh \left[\frac{\Phi(\Bsigma)}{N}\right].
\end{equation}

The maximal work that can be extracted from a cycle, in this case equals the one that can be extracted from a Szil\'ard box with $q=2$ partitions and $N$ particles (see Fig. \ref{fig:szexamples} and Eq. \eqref{UC_SZ} for $N\gg 1$). 

Analogous considerations can be applied to the model 
\begin{equation}
\label{spinh}
P_{\rm eq}(\Bsigma|h)=\frac{e^{h \mathbf{M}^{z}(\Bsigma)}}{(4\cosh h)^N},\qquad \mathbf{M}^{x}(\Bsigma)=\sum_{i=1}^N\sigma_i^{x},
\end{equation}
where again the MLE value is given by $\hat h(\Bsigma)={\rm arg}\tanh [\mathbf{M}^{x}(\Bsigma)/N]$. The partitions are shown in Fig. \ref{figExample}(e) for this case, as a function of $h$. Notice that the two cases differ only by a permutation of the partitions, hence the value of $\mathcal{R}^*$ is the same as above.

In a system where both  $\mathbf{M}^{x}$ and $\BPhi$ can be measured, the optimal non-equilibrium state to which the system should be driven would be given by the NML of the model 
\begin{equation}
\label{spinJh}
P_{\rm eq}(\Bsigma|J,h)=\frac{e^{h \mathbf{M}^{z}(\Bsigma)+J\Phi(\Bsigma)}}{(16\cosh h\cosh J)^N},\qquad \mathbf{M}^{x}(\Bsigma)=\sum_{i=1}^N\sigma_i^{x}.
\end{equation}
The energy levels $\epsilon_{\vec{\sigma}}(h,J)$ depend on both $h$ and $J$, which provides more flexibility in adjusting partitions. Hence more work can be extracted. Detailed calculation shows that, in this case, twice as much work can be extracted, compared to the previous cases.

\section{Discussion}
\label{sec:iv}

The result of this paper lends itself to a number of considerations: 

\subsection{Efficient communication as ideal information engines}

The process of measuring a quantity $\Phi$ can be seen as a communication problem between two parties, Alice 
 and Bob. 
 In order to focus on thermodynamic efficiency, we model the channel over which Alice and Bob communicate, as an information engine. The messages that Alice and Bob exchange are a sequence $\mathbf{s}=(s_1,\ldots,s_N)$ of symbols from a finite alphabet, where $s_i$ are drawn i.i.d. from a certain distribution.  The exchange is realized by an information engine that operates at a constant temperature. Let $\Phi(\mathbf{s})$ be the value of the quantity that is measured. Alice and Bob both have a device, that is initially in state $P_0(\mathbf{s})$. They can choose to code their messages with a distribution $P_{\rm neq}(\mathbf{s})$. This requires Alice to spend $\Delta F=D[P_{\rm neq}(\mathbf{s})||P_0(\mathbf{s})]$ units of $k_BT$ to generate messages, but it allows Bob to harvest  work $-W(\Phi)$ from the measurement of $\Phi(\mathbf{s})$. If Alice and Bob exchange messages both ways it makes sense for them to agree on a design 
 that minimises the average energetic cost $\Delta F+\langle W\rangle$ of the communication, over many cycles. Our result implies that {\rm (i)} $P_0$ should be of the form Eq. (\ref{model1}); {\rm (ii)} $P_{\rm neq}=P^*$ should be the corresponding NML, and {\rm (iii)} some work can be extracted for each message, that on average cannot exceed the minimax redundancy $\mathcal{R}^{\star}$ times $k_BT$. \MM{The channel should work as follows: A value of $\BPhi$ is drawn at random from $P^*(\BPhi)$. Knowing $\BPhi$, Alice can prepare the engine with partitions of sizes dictated by $\hat\Btheta(\BPhi)$ and distribute the $N$ particles in such a way that $\BPhi(\bs)$ takes the desired value $\BPhi$. Then she can operate a quasistatic transformation to bring back the partitions to the sizes that correspond to the equilibrium values $\Btheta$, spending $\Delta F$ units of $k_BT$. Then Alice can pass the information engine to Bob, who would measure $\BPhi$ and extract work from the isothermal expansion of the partitions.} 
 
  Interestingly, the costs and benefits across the communication channel are shared unequally in this optimal protocol.
  This argument neglects the cost needed to erase the memory in which $\BPhi$ is stored, which is at least equal to the Shannon entropy $H[\BPhi]$ of $\BPhi$. This makes sense in a communication setting, where $\BPhi$ is valuable information that does not need to be erased. \MM{In addition, we show that this communication protocol provides an energy budget that can cover half of the memory cost, in the limit of long messages.}  
  

\subsection{The importance of being critical}

The NML codes have been recently shown to be ``critical'' in a very precise sense. First, Ref.~\cite{CuberoMDL} has shown that, a typical sample drawn from $P^*$ exhibits {\em statistical criticality} \cite{Mora}, i.e. the distribution of $n_s$ is very broad. Second, studying large deviations of the in-sample description length per character, $\hat H[s]=-\sum_{s=1}^q \hat\ell_s\ln\hat \ell_s$, for samples generated from $P^*$, Ref. \cite{CuberoMDL} has shown that $P^*$ sits exactly at a critical point. This is because there are no codes that can achieve a better compression than $P^*$. It is interesting that extracting the maximal amount of work from the cycle of an information engine, requires driving the system to a critical state. This provides hints for possible experimental protocols towards more efficient information engines. 

\subsection{Which observables should be measured?}

The above discussion suggests that most efficient communication is achieved for sufficient statistics  $\Phi$ such that $\mathcal{R}^{\star}$ is largest. This result provides further insights with respect to recent work~\cite{matsumoto2018role} which pointed to the relevance of  the existence of sufficient statistics in information-thermodynamic efficiency.  We have shown that, for fixed number of particles $N$ and parameters $M$, the maximal work extracted by a broad class of information engines depends on the value $c=\int \text{d}\theta\sqrt{\det{\mathbf{J}(\theta)}}$ of the constant term in Eq. (\ref{cexp}), which is reminiscent of the so-called "thermodynamic length" introduced in Refs.~\cite{sivak2012thermodynamic,crooks2007measuring,feng2009far}. Observables for which the corresponding value of $c$ is larger should allow for more energy efficient measurement devices. For example Ref.~\cite{beretta2018stochastic} has studied the term $ c$ for spin models with interactions of arbitrary order and found that pairwise models achieve large values of $c$.

%
%

\subsection{Experimental applications}


Our results shed light on potential future experimental applications in stochastic thermodynamics. Until now, experimental realizations of one-particle Szil\'ard engines have been achieved using e.g. colloidal particles trapped with optical tweezers~\cite{toyabe2010experimental,berut2012experimental,roldan2014universal} and feedback traps~\cite{exp-landauer2014}, DNA hairpins~\cite{ribezzi2019large}, and single-electron transistors~\cite{koski2014experimental}. Most of these experiments rely on the accurate control of a single degree of freedom, e.g. the position of a colloid along the $x$ axis, the energy level occupied by one extra electron in a single-electron transistor, etc. This leads to optimal work extractions of the order of ~$k_{\rm B}T$ by means of current experimental techniques. 

We have shown that Szil\'ard engines extracting work above the "$k_{\rm B}T$ limit" can be realized physically, e.g. with $N$-particle  systems in $q$ energy levels. Our formalism provides exact formulas for the non-equilibrium state at which an $N$-particle information engine should be driven and how much work can be extracted from the measurement of an arbitrary coarse-grained physical quantity. The potential theoretical extensions to bosonic systems and experimental realizations of these results with e.g. polaritons or photons trapped in optical cavity networks~\cite{reiserer2015cavity,hartmann2006strongly}, are two interesting avenues of future research.

\section{Acknowledgments}

L.T. thanks The Abdus Salam International Centre for Theoretical Physics (ICTP) for hospitality. 
We thank Antonio Celani, Sebastian Goldt, Gonzalo Manzano, Juan M.~R. Parrondo, Isaac Perez-Castillo, Diego Porras and Susanne Still for stimulating discussions and comments.

\appendix

\section{The entropy of the measurement}
\label{appH}

Let us compute the entropy of measurement in the case of a generic sufficient statistics $\BPhi$ and then specialise to the case of the standard Szil\'ard engine, where $\BPhi=\bn$. For finite $N$, $\BPhi$ takes values in a discrete set. 
The entropy of $\BPhi$ is defined as:
\begin{equation}
H[\BPhi]=-\sum_{\BPhi}P^*(\BPhi)\ln P^*(\BPhi)
\end{equation}
where 
\begin{equation}
P^*(\BPhi)=\sum_{\bs}P^*(\bs)\delta\left(\BPhi-\BPhi(\bs)\right)=e^{-\cR}\sum_{\bs}P(\bs|\hat\Btheta(\bs))\delta\left(\BPhi-\BPhi(\bs)\right).
\label{eqPPhi}
\end{equation}
For large $N\gg M$ we can use the relation 
\begin{equation}
P(\bs|\hat\Btheta(\bs))\simeq \left(\frac{N}{2\pi}\right)^{M/2}\int\! d\Btheta \sqrt{{\rm det}J(\Btheta)}P(\bs|\Btheta)
\end{equation}
which is easily verified by performing the integral on the right hand side by the saddle point method (see e.g. \cite{CuberoMDL}). 
Here and below the $\simeq $ sign stands for an asymptotic equality, that holds in the limit $N\to\infty$, in the sense that terms that are neglected on the right hand side vanish as $N\to\infty$.
Using also 
\begin{equation}
\cR\simeq\frac M 2 \ln N+c,\qquad c\equiv \ln \int\! d\Btheta \sqrt{{\rm det}J(\Btheta)},\label{cexp1}
\end{equation}
we rewrite Eq. (\ref{eqPPhi}) as
\begin{equation}
P^*(\BPhi)\simeq e^{-c}\int\! d\Btheta \sqrt{{\rm det}J(\Btheta)}P(\BPhi|\Btheta)
\end{equation}
with 
\begin{equation}
\label{eqexp}
P(\BPhi|\Btheta)=\sum_{\bs}
P(\bs|\Btheta)\delta\left(\BPhi-\BPhi(\bs)\right)=e^{\Btheta\BPhi+\cF_{\rm eq}(\Btheta)+\cS(\BPhi)},
\end{equation}
where 
\begin{equation}
\cS(\BPhi)=\ln \sum_{\bs}\delta\left(\BPhi-\BPhi(\bs)\right)
\end{equation}
is the entropy (i.e. the log of the number) of configurations with $\BPhi(s)=\BPhi$. All quantities in the exponential of Eq. (\ref{eqexp}) are extensive. Let us introduce to the intensive variable 
\begin{equation}
\bar\Bphi=\frac \BPhi N.
\end{equation}
For $N$ large, $\bar\Bphi$ takes values on a dense set, so we change sums over $\BPhi$ into integrals over $\bar\Bphi$
\begin{equation}
\sum_{\BPhi}(\ldots)\simeq {N^M} \int\! d\bar\Bphi (\ldots),
\end{equation}
where the factor $N^M$ comes from the Jacobian of the transformation $\BPhi\to\bar\Bphi$. 
Correspondingly, distributions change into probability density functions, as
\begin{equation}
P(\BPhi|\Btheta) \simeq N^{-M}P(\bar\Bphi|\Btheta), ~~~ P^*(\BPhi)  \simeq N^{-M}p^*(\bar\Bphi).
\end{equation}
We also introduce the intensive functions
\begin{equation}
\varphi(\Btheta)=\frac 1 N \cF_{\rm eq}(\Btheta),\qquad\sigma(\bar\Bphi)=\frac 1 N \cS(N\bar\Bphi).
\end{equation}
With this change of variables, the entropy now reads:
\begin{equation}
H[\BPhi]\simeq M\ln N-\int d\bar\Bphi\, p^*(\bar\Bphi)\ln p^*(\bar\Bphi)
\end{equation}
Now, $p(\bar\Bphi|\Btheta)$ is a sharply peaked function on the value $\hat\Bphi(\theta)$ that dominates the saddle point, i.e. such that $\sigma'(\bar\Bphi)=\Btheta$. Notice that $-\sigma(\bar\Bphi)$ is the Legendre transform of $\varphi(\Btheta)$ and vice-versa.
In the limit $N\to\infty$ we can write
\begin{equation}
p(\bar\Bphi|\Btheta)\simeq \delta\left(\bar\Bphi-\hat\Bphi(\Btheta)\right)
\end{equation}
Therefore
\begin{eqnarray}
p^*(\bar\Bphi) & \simeq & e^{-c}\int\! d\Btheta\sqrt{{\rm det}J(\Btheta)}\delta\left(\bar\Bphi-\hat\Bphi(\Btheta)\right)\label{eqdphi} \\
 & = & e^{-c}\int \! d\Btheta\sqrt{{\rm det}J(\Btheta)}\frac{\delta\left(\Btheta-\hat\Btheta(\bar\Bphi)\right)}{{\rm det}J(\Btheta)}\\
 & = & e^{-c}\frac{1}{\sqrt{{\rm det}J(\hat\Btheta(\bar\Bphi))}}\\
 & = & e^{-c}\sqrt{{\rm det}K(\bar\Bphi)}\label{pstardens}
\end{eqnarray}
where, in Eq. (\ref{eqdphi}), we changed variables $\bar\Bphi\to\Btheta$ in the delta function, using the fact that the Jacobian is given by 
\begin{equation}
J(\Btheta)=\frac{d\hat\Bphi}{d\Btheta},
\end{equation}
which is the Fisher Information matrix. 
The matrix $K$ is instead the Jacobian of the inverse transformation $\Btheta\to\bar\Bphi$ 
\begin{equation}
K(\bar\Bphi)=\frac{d\hat\Btheta(\bar\Bphi)}{d\bar\Bphi}=J^{-1}\left(\hat\Btheta(\bar\Bphi)\right).
\end{equation}
Note that Eq. (\ref{pstardens}) is correctly normalised, because, changing variables $\Btheta\to\bar\Bphi$  in Eq. \ref{cexp1}), one finds
\begin{eqnarray}
c & = & 
\ln\int \! d\Btheta \sqrt{{\rm det}J(\Btheta)}\\
& = & 
\ln\int \! d\bar\Bphi \, {\rm det}K(\bar\Bphi) \sqrt{{\rm det}J\left(\hat\Btheta(\bar\Bphi)\right)}\\
& = & \ln\int\!d\bar\Bphi \sqrt{{\rm det}K(\bar\Bphi)}.
\end{eqnarray}
Therefore in the limit $N\to\infty$ 
\begin{eqnarray}
H[\BPhi] & \simeq & M\ln N+c-e^{-c}\int d\bar\Bphi  \sqrt{{\rm det}K(\bar\Bphi)}\ln \sqrt{{\rm det}K(\bar\Bphi)} \\
 & = & M\ln N+c+e^{-c}\int d\Btheta \sqrt{{\rm det}J(\Btheta)} \ln \sqrt{{\rm det}J(\Btheta)} \\
 & = & M\ln N+\ln\int \! d\Btheta \sqrt{{\rm det}J(\Btheta)}+\frac{\int d\Btheta \sqrt{{\rm det}J(\Btheta)} \ln \sqrt{{\rm det}J(\Btheta)}}{\int \! d\Btheta \sqrt{{\rm det}J(\Btheta)}}
 \label{last}
\end{eqnarray}
which, together with Eq. (\ref{cexp1}), implies that
\begin{equation}
\lim_{N\to\infty}\frac{\cR}{H[\BPhi]}=\frac 1 2.
\end{equation}
In the case of the Szil\'ard engine with $\BPhi=\bn$, we have (see e.g. \cite{beretta2018stochastic})
\begin{equation}
{\rm det}J(\Bell)=\prod_{s=1}^q\ell_s^{-1}\delta\left(\sum_{s=1}^q\ell_s-1\right).
\end{equation}
We can use the identity
\begin{equation}
\int_0^1\! d\Bell \prod_{s=1}^q\ell_s^{\alpha-1}\delta\left(\sum_{s=1}^q\ell_s-1\right)=\frac{\Gamma(\alpha)^q}{\Gamma(q\alpha)}\equiv e^{\Upsilon(\alpha)},
\end{equation}
which also defines the function $\Upsilon(\alpha)$. Upon inspection, we find that $c=\Upsilon(1/2)$ and the constant term in Eq. (\ref{last}) is given by
\[
-\left.\frac{d\Upsilon}{d\alpha}\right|_{\alpha=1/2}=q\left[\psi(q/2)-\psi(1/2)\right],
\]
as in Eq. (\ref{Hnasymp}). 

\section{A physical realisation with $q=4$ partitions}
\label{appspin}

Let us consider as an example N classical particles in a 2D plane, with coordinates $\mathbf{r}_i=(x_i,y_i)$, 
confined by a potential $V(r)$, centered at $\mathbf{r}=0$ and invariant under rotation around the origin (here $r=|\mathbf{r}|=\sqrt{x^2+y^2}$ denotes the distance from the origin). We assume that $k_BT=1$ for simplicity.
We label each particle with a spin representation $s=\vec{\sigma}=(\sigma^{x},\sigma^{y})$, with $\sigma^{x}={\rm sign}(x)$ and $\sigma^{y}={\rm sign}(y)$ that identify the quadrant to which the point $\mathbf{r}=(x,y)$ belongs. 

The distribution of the positions of the particles in absence of the walls is:
\begin{eqnarray}
P_{\rm eq}(\mathbf{r}_1,...,\mathbf{r}_N)&=&\frac{e^{-\sum_i V({r}_i)}}{Z}\\
Z&=&\left(\int\!d^2\mathbf{r}e^{-V({r})}\right)^N=\left(2\pi \int_0^\infty\! rd{r}e^{-V({r})}\right)^N
\end{eqnarray}

When the walls are inserted along the axes, the particles with different spin labels are separated into four partitions, that correspond to the four quadrants. The walls can rotate around the origin in order to extract work, with the labels of particles remaining unchanged. We denote by $n_{\vec{\sigma}}=\sum_{i=1}^N \delta_{\sigma_i^x,\sigma^x} \delta_{\sigma_i^y,\sigma^y}$ the number of particles with label $\vec{\sigma}=(\sigma^{x},\sigma^{y})$.

%

When the walls are moved so that the two walls delimiting the domain of particles with label $\vec{\sigma}$ make an angle $\alpha_{\vec{\sigma}}$, the new partition function is given by:
\begin{equation}
Z(\vec{\alpha})=\prod_{\vec{\sigma}}Z_{\vec{\sigma}}(\alpha_{\vec{\sigma}},n_{\vec{\sigma}})
\end{equation}
with 
\begin{equation}
    Z_{\vec{\sigma}}(\alpha_{\vec{\sigma}},n_{\vec{\sigma}})=\left(\int_0^{\alpha_{\vec{\sigma}}}d\theta\int_0^{+\infty}dre^{-V(r)}\right)^{n_{\vec{\sigma}}}=\alpha_{\vec{\sigma}}^{n_{\vec{\sigma}}}\left(\int_0^{+\infty}dre^{-V(r)}\right)^{n_{\vec{\sigma}}}.
\end{equation}
The free energy is $F(\vec{\alpha})=-\ln Z(\vec{\alpha})$, so the pressure in partition $\vec{\sigma}$ is given by:
\begin{equation}
    \rm \mathsf{P}_{\vec{\sigma}}(\alpha_{\vec{\sigma}},n_{\vec{\sigma}})=-\frac{\partial F({\vec{\alpha}})}{\partial\alpha_{\vec{\sigma}}}=\frac{n_{\vec{\sigma}}}{\alpha_{\vec{\sigma}}}
\end{equation}

In order to extract maximum work, one should move the walls so that the pressure is the same in the four partitions.
This implies
\begin{equation}
\alpha_{\vec{\sigma}}=2\pi\frac{n_{\vec{\sigma}}}{N}.
\end{equation}
This system is then very similar to a Szil\'ard box with $q=4$. Indeed in terms of the variable $\ell_{\vec{\sigma}}=\frac{\alpha_{\vec{\sigma}}}{2\pi}$, the equilibrium distribution is given exactly by Eq. (\ref{model}). 

The model in Eq. (\ref{spinJ}) is realised in a system where the two vertical walls are constrained to rotate together as if they formed a single wall, and that the horizontal walls are fixed. In order to see this, we call $\alpha$ the angle that the vertical wall forms with the $x$-axis (i.e. with the fixed wall). Then:
\begin{equation}
\label{spinalphaJ}
    P_{\rm eq}(\boldsymbol{\sigma}|\alpha)=\frac{1}{(2\pi)^N}\alpha^{n_{++}+n_{--}}(\pi-\alpha)^{n_{+-}+n_{-+}}
\end{equation}
This coincides with Eq. (\ref{spinJ}), upon defining $J=-\frac{1}{2}\ln(\frac{\pi}{\alpha}-1)$, so that $\frac{e^{J}}{4\cosh J}=\frac{\alpha}{2\pi}$ and $\frac{e^{-J}}{4\cosh J}=\frac{\pi-\alpha}{2\pi}$. The same argument shows that the model in Eq. (\ref{spinh}) corresponds to the case where the two vertical walls are constrained to rotate in opposite directions, while the two horizontal walls stay fixed. 

\bibliographystyle{ieeetr}
\bibliography{SZE_small}

\end{document}